\documentstyle[epsfig]{elsart}

\def\bq{\begin{equation}}
\def\eq{\end{equation}}
\def\be{\begin{eqnarray}}
\def\ee{\end{eqnarray}}
\def\ci{\cite}
\def\bi{\bibitem}

\def\eep{(e,e^\prime p)}
\def\ptp{(p,2p)}
\def\PkE{P(k,E)}
\def\primo{^\prime}
\def\lsim{\mathrel{\rlap{\lower4pt\hbox{\hskip1pt$\sim$}}
    \raise1pt\hbox{$<$}}}         %less than or approx. symbol
\def\gsim{\mathrel{\rlap{\lower4pt\hbox{\hskip1pt$\sim$}}
    \raise1pt\hbox{$>$}}}         %greater than or approx. symbol

\begin{document}

\begin{frontmatter}

\title{Final state interactions in $^{4}$He(e,e$^\prime$p)$^{3}$H at 
large proton energy}
   
\author[Rome]{O. Benhar}
\author[Julich,Moscow]{N.N. Nikolaev}
\author[Julich]{J. Speth}
\author[Aligarh]{A.A. Usmani}
\author[Moscow]{B.G. Zakharov}

\address[Rome]{INFN, Sezione Roma 1, I-00185 Rome, Italy}
\address[Julich]{IKP(Theorie), Forschungszentrum  J\"ulich GmbH.
 D-52425 J\"ulich, Germany}
\address[Moscow]{L.D.Landau Institute for Theoretical Physics, 
117940,~ul.~Kosygina~2,~V-334~Moscow, Russia}
\address[Aligarh]{Physics Department, Aligarh Muslim University, 
Aligarh-202 002, India}

\begin{abstract}

At large proton energy, the final state interactions of the knocked out nucleon
 in $\eep$ reactions off nuclear targets can be described within the eikonal
approximation, treating the spectator particles as a collection of fixed
scattering centers. We use a generalization of this approach, suitable to 
take into account the possible occurrence of color transparency, to
carry out an accurate calculation of the missing momentum distribution of the 
process $^{4}He(e,e^\prime p)^{3}H$. The pattern of final state interaction
effect is analyzed for different kinematical setups in the domain corresponding
to $2 \le Q^2 \le 20$ (GeV/c)$^2$.

\end{abstract}


\end{frontmatter}

%%%%%%%%%%%%%%%%%%%%%%%%%%%
\section{Introduction}
%%%%%%%%%%%%%%%%%%%%%%%%%%%

Electron-nucleus scattering experiments in which a proton is detected
in coincidence with the outgoing electron have long been recognized as
a powerful tool to study both nuclear and nucleon dynamics 
(see, e.g., ref.\ci{book}). According
to the Plane Wave Impulse Approximation (PWIA), which is
expected to be valid at large momentum transfer, the nuclear
$\eep$ cross section reduces to the incoherent sum of the
cross sections off individual nucleons, whose distribution in momentum
${\bf k}$ and
removal energy $E$ is dictated by the spectral function $\PkE$, the
final state interactions (FSI) between the knocked out particle and the
recoiling spectator system being negligible. As a consequence, the PWIA
cross section of the process in which an electron of initial energy $E_i$ is
scattered into the solid angle $\Omega_e$ with energy $E_f=E_i-\omega$,
while a proton of kinetic energy $T_p$ is ejected into the solid angle $\Omega_p$,
takes the simple factorized form
\bq
\frac{d\sigma}{d\omega d\Omega_e d\Omega_p dT_p}
= p(T_p+m) {\widetilde \sigma}_{ep} P(p_m,E_m)\ ,
\label{xsec}
\eq
where $m$ denotes the nucleon mass, while the missing momentum ${\bf p}_m$ and
missing energy $E_m$ are defined as
\bq
{\bf p}_m  = {\bf p}-{\bf q}
\label{miss:mom} \\
\eq
and
\bq
E_m  =  \omega-T_p-T_R\ .
\label{miss:en}
\eq
In the above equations, ${\bf q}$ is the momentum transfer and
$T_R=p_R^2/M_{A-1}$, with ${\bf p}_R=-{\bf p}_m$, is the kinetic
energy of the recoiling spectator system of mass $M_{A-1}$. The cross section
${\widetilde \sigma}_{ep}$ of eq.(1) describes
electron scattering off a {\it bound} nucleon of momentum
${\bf p}_m$ and removal energy $E_m$ \ci{defo}.

In presence of nonnegligible FSI, the PWIA picture breaks down, and the missing 
momentum and energy cannot be readily interpreted as the initial momentum and 
removal energy of the outgoing nucleon. Therefore, 
a quantitative understanding of FSI is 
 needed in order to extract the information on the nucleon spectral 
function from the measured $\eep$ cross section. A wealth of highly accurate 
theoretical calculations of FSI effects in $\eep$ reactions have been 
carried out within the
Distorted Wave Impulse Approximation (DWIA), in which the interaction between
the knocked out nucleon and the spectator system is described in terms of a 
complex optical potential (see, e.g., ref.\ci{DWIA}). Using the results of these
calculations it has been possible to obtain the spectral functions
 describing the single-particle states, predicted by the nuclear shell 
model, from the analysis of the available low missing energy data \ci{pka}.

It has to be emphasized, however, that FSI should not only be regarded as 
a {\it noise}, 
to be removed from the measured cross sections. In fact, in many instances FSI 
produce a {\it signal} that carries relevant information on both the 
 target structure and the dynamics of the scattering process. For example, it 
has been shown that FSI effets, which obviously depend upon the distribution in 
space of the spectator particles, are very sensitive to the presence of 
local fluctuations of the nuclear density produced by nucleon-nucleon (NN) 
correlations \ci{bho}. 

The analysis of FSI in $\eep$ processes may also provide information 
on NN scattering in the nuclear medium. At 
moderate proton energies ($\sim$ 100 MeV) {\it nuclear} structure effects, 
such as Pauli blocking and dispersive corrections, lead to significant changes 
in the NN scattering amplitude \ci{papi}. While these effects are expected to 
become negligible for proton energies in the few GeV range, different 
effects, arising from {\it nucleon} structure, may become important in this 
kinematical regime, corresponding to high $Q^2$ ($Q^2 = q^2 - \omega^2$). 

Perturbative Quantum Chromo-Dynamics (QCD) predicts that
elastic scattering on a nucleon at high momentum transfer can only occur
if the nucleon is found in the Fock state having the lowest number of 
constituents, so that the momentum can be most effectively shared among them.
 This state, being very compact, interacts weakly with the spectator
particles and evolves to the standard proton configuration with a 
characteristic timescale that increases with the momentum transfer. According 
to this picture a proton, after absorbing a large momentum $q$, e.g. in 
an electron
scattering process, can travel through the spectator system experiencing 
very little attenuation, i.e. exhibits  {\it color transparency} 
(CT) \cite{CT1,CT2}. In the limit $Q^2 \rightarrow \infty$ 
FSI effects in $\eep$ are expected to become vanishingly small.

The possible signatures of the occurrence of CT in coincidence $\eep$ and 
$\ptp$ processes have been recently studied within a theoretical many-body 
approach suitable for the calculation of semi-inclusive 
cross sections, involving a sum over the states of the undetected 
spectator system \ci{asy,eep,ptp}. The treatment of FSI of 
refs.\ci{asy,eep,ptp} is based on a generalization of Glauber theory 
of high energy proton scattering \ci{glauber}. 

In this paper we extend the approach of refs.\ci{asy,eep,ptp} to the case 
of fully exclusive reactions, in which the final state of the recoiling nucleus 
is specified. Our treatment of the corresponding amplitude is presented 
in section II, where we discuss 
both the many-body aspects, related to the description of the nuclear 
initial and final states, and the structure of the scattering 
operator, modeling the FSI of the knocked out proton and the transition to the CT 
regime. The results obtained applying our approach to the case of 
a $^{4}$He target, in which accurate numerical calculations are feasible, are given 
in section III, where FSI effects on different observables are discussed.
Finally, the summary and conclusions are presented in section IV.

%%%%%%%%%%%%%%%%%%%%%%%%%%%%%%%%%%%%%%%%%%%%%%%%
\section{Formalism}
%%%%%%%%%%%%%%%%%%%%%%%%%%%%%%%%%%%%%%%%%%%%%%
\subsection{\boldmath $(e,e^\prime p)$ \unboldmath  amplitude at high 
proton energy}

We will focus on $\eep$ processes in which the recoiling (A-1)-particle system
is left in a bound state $|\varphi_n \rangle$. Neglecting 
many-body contributions to the electromagnetic current, the nuclear matrix 
element associated with the transition amplitude can be written
\bq
M_n({\bf p},{\bf q}) = \langle \Psi^{(-)}_{n {\bf p}} |
   \sum_{\bf k} a_{{\bf k}+{\bf q}}^\dagger a_{\bf k} | \Psi_0 \rangle\ ,
\label{mat:el}
\eq
where $a_{{\bf k}+{\bf q}}^\dagger$ ($ a_{\bf k}$) denotes the usual 
creation (anihilation) operator and the target ground state 
$| \Psi_0 \rangle$ satisfies the Schr\"odinger equation 
$ H_A | \Psi_0 \rangle = E_0 | \Psi_0 \rangle\ $.

The terms responsible for FSI can be isolated in $H_A$ rewriting the nuclear
hamiltonian in the form
\bq 
H_A = \sum_{i=1}^A T_i + \sum_{j>i=1}^A v_{ij} = H_0 + H_1\ ,
\label{decomp}
\eq
with (the knocked out nucleon is labelled with index 1)
\bq
H_0 = \sum_{i=1}^A T_i + \sum_{j>i=2}^A v_{ij} = H_{A-1} + T_1\ ,
\eq
and
\bq
H_1 = \sum_{j=2}^A v_{1j}\ .
\eq
In the above equations $T_i$ and $v_{ij}$ denote the kinetic energy of the 
$i$-th nucleon and the interaction potential between nucleons $i$ and $j$, 
respectively. $H_0$ is the PWIA hamiltonian, describing the system containing 
(A-1) interacting spectators and the noninteracting knocked out nucleon, whereas 
the terms associated with FSI are included in $H_1$.

The decomposition of eq.(\ref{decomp}) can be used to write the  
final scattering state $| \Psi^{(-)}_{n {\bf p}} \rangle$, 
in the form \cite{gold}:
\bq
| \Psi^{(-)}_{n {\bf p}} \rangle = \Omega^{(-)}_{{\bf p}}
  | \Phi_{n {\bf p}} \rangle \ ,
\label{fin:wf}
\eq
where $| \Phi_{n {\bf p}} \rangle$ denotes the asymptotic state 
with no interaction between particle $1$ and the spectators, which is obviously
an eigenstate of $H_0$. In coordinate space it can be written 
\bq
\Phi_{n {\bf p}} (R) = \sqrt{\frac{1}{V}}\  
e^{i{\bf p} \cdot {\bf r}_1} \varphi_n({\widetilde R})\ ,
\label{asy:wf}
\eq
where $V$ is the normalization volume, while 
$R \equiv \{ {\bf r}_1, {\bf r}_2 \ldots ,{\bf r}_A \}$ 
and ${\widetilde R} \equiv \{ {\bf r}_2, \ldots ,{\bf r}_A \}$ specify
the configurations of the full A-particle system and the (A-1)-particle 
spectator system, respectively. 

Setting $\Omega^{(-)}_{{\bf p}}=1$, which amounts to disregarding the 
effects of FSI, and substituting into eq.(\ref{mat:el}), one obtains the PWIA 
amplitude, depending upon the the missing momentum ${\bf p}_m = {\bf p}-{\bf q}$ 
only. The operator $\Omega^{(-)}_{{\bf p}}$ describes the distortion of the 
asymptotyc wave function produced by the rescattering of the 
knocked out nucleon. It can be formally written as
\bq
\Omega^{(-)}_{{\bf p}} = \lim_{t \rightarrow \infty} e^{iH_At}
e^{-iH_0t} =  \lim_{t \rightarrow \infty} {\widehat T} 
{\rm e}^{-i \int_0^t dt\primo\ {\widehat H}_1(t\primo)}\ ,
\label{omega:def}
\eq
where ${\widehat T}$ denotes the time ordering operator and 
\bq
{\widehat H}_1(t) ={\rm e}^{iH_0t}H_1{\rm e}^{-iH_0t}\ . 
\eq

In general, the calculation of $\Omega^{(-)}_{{\bf p}}$ from 
eq.(\ref{omega:def}) with a realistic
nuclear hamiltonian involves prohibitive difficulties. However, when
the kinetic energy carried by the knocked out proton is large, the
structure of $\Omega^{(-)}_{{\bf p}}$ can be strongly simplified using 
 a generalization of the approximation scheme originally developed by Glauber 
to decribe 
proton-nucleus scattering \ci{glauber}.
The basic assumptions underlying this scheme are that 
i) the fast struck nucleon moves along a straight trajectory, being 
undeflected by rescattering processes ({\it eikonal approximation}) and 
ii) the spectator system can be seen as a collection of fixed scattering 
centers ({\it frozen approximation}).

Implementation of the eikonal and frozen approximations in 
the definition of the scattering operator 
$\Omega^{(-)}_{{\bf p}}$, eq.(\ref{omega:def}), leads to the following 
coordinate space expression:
\be
\nonumber
\Omega^{(-)}_{{\bf p}}(R) & = & \langle R |\Omega^{(-)}_{{\bf p}}| R \rangle
= P_z\ \prod_{j=2}^A \left[ 1 - \Gamma_{p}(1,j) \right]\\
& = & P_z\ \left[ 1 - \sum_{j=2}^A \Gamma_{p}(1,j) + 
\sum_{k>j=2}^A \Gamma_{p}(1,j)
\Gamma_{p}(1,k) - \ldots \right] \ ,
\label{omega:hea}
\ee
where the positive $z$-axis is chosen along the eikonal trajectory and the 
$z$-ordering operator $P_z$ prevents the occurrence of backward 
scattering of the fast struck proton. The structure of the two-body operator 
$\Gamma_{p}(1,j)$, describing the dynamics 
of the scattering process, will be discussed in the next section.

Inserting $\Omega^{(-)}_{{\bf p}}$ of eq.(\ref{omega:hea}) into
the definition of $| \Psi^{(-)}_{n {\bf p}} \rangle$ 
of eq.(\ref{fin:wf}) one gets the following 
expression for the matrix element of eq.(\ref{mat:el}):
\bq
M_n({\bf p},{\bf q}) = \int dR\ \left[ 
\varphi_n({\widetilde R})\Omega^{(-)}_{{\bf p}}(R) \right]^\ast
 {\rm e}^{i({\bf p}-{\bf q}) \cdot {\bf r}_1} \Psi_0(R)\ .
\label{matel:hea}
\eq

The calculation of the above amplitude can be simplified introducing a further
approximation, whose validity rests on the same assumptions made to justify
the use of the frozen approximation. Within this scheme \ci{elba}, one replaces the 
many-body scattering operator $\Omega^{(-)}_{{\bf p}}(R)$ with a one-body operator, 
depending on the position of the knocked out nucleon only, that can be obtained
averaging $\Omega^{(-)}_{{\bf p}}(R)$ over the positions of the spectator particles 
in the target ground state according to the following definition:
\bq
{\overline \Omega}^{(-)}_{{\bf p}}({\bf r}) =
\frac{1}{\rho_A({\bf r})}
\int dR\ |\Psi_0(R)|^2\ \Omega^{(-)}_{{\bf p}}(R)\
\frac{1}{A} \sum_{i=1}^A \delta({\bf r}-{\bf r}_i)\ ,
\label{average:op}
\eq
where $\rho_A({\bf r})$ is the target density normalized to unity.

Substitution of $\Omega^{(-)}_{{\bf p}}(R)$ with 
${\overline \Omega}^{(-)}_{{\bf p}}({\bf r})$ in eq.(\ref{matel:hea}) 
allows one to rewrite the amplitude in the form:
\bq
M_n({\bf p},{\bf q}) = \int d^3r\
 {\rm e}^{i({\bf p}-{\bf q}) \cdot {\bf r}} \psi_{n {\bf p}}({\bf r})\ ,
\label{addition}
\eq
the {\it distorted  overlap} $\psi_{n {\bf p}}({\bf r})$ being defined as:
\bq
\psi_{n {\bf p}}({\bf r}) =
\left[ {\overline \Omega}^{(-)}_{{\bf p}}({\bf r}) \right]^\ast
 \chi_{n}({\bf r})\ ,
\label{dist:qh}
\eq
with 
\bq
\chi_{n}({\bf r}_1)=\int d{\widetilde R}\ 
\varphi_n^\ast({\widetilde R}) \Psi_0(R)\ .
\label{quasih}
\eq
Note that, within the nuclear shell model picture,  
the quantity defined in eq.(\ref{quasih}) can be interpreted as the 
wave function associated with the single particle state 
initially occupied by the knocked out nucleon \cite{rbpp}.

The overlap relevant to the case of proton knock out from a $^{4}$He target
leading to a recoiling $^3$H can be written:
\bq
\chi_0({\bf X}) = \int d^3Y d^3Z \ 
\Psi_3^\ast({\bf Y},{\bf Z})\ \Psi_4({\bf X},{\bf Y},{\bf Z})\ ,
\label{chi:zero}
\eq
with ${\bf Y}={\bf r}_2-{\bf r}_3$, 
${\bf Z}=(2/3){\bf r}_4 -({\bf r}_2+{\bf r}_3)/3$, 
${\bf X}={\bf r}_1-({\bf r}_2+{\bf r}_3+{\bf r}_4)/3$, whereas 
$\Psi_3({\bf Y},{\bf Z})$ and $\Psi_4({\bf X},{\bf Y},{\bf Z})$ denote the 
ground state wave functions of $^{3}$H and $^{4}$He, respectively. The 
function 
$\chi_0({\bf X})$ has been evaluated by Schiavilla {\it et al} with highly 
realistic wave functions, obtained using the Variational Monte Carlo approach 
and nuclear hamiltonians including two- and three-nucleon interactions \ci{SPW}. 
We have used 
the results of ref.\ci{SPW} to calculate the amplitude of eq.(\ref{addition})  
with the averaged scattering operator given by eq.(\ref{average:op}), 
whose definition in the $^4$He center
of mass frame reads
\bq
{\overline \Omega}^{(-)}_{{\bf p}}({\bf X}) = \frac{
\int d^3Y d^3Z\ |\Psi_4({\bf X},{\bf Y},{\bf Z})|^2\ 
\Omega^{(-)}_{{\bf p}}({\bf X},{\bf Y},{\bf Z}) }
{\int d^3Y d^3Z\ |\Psi_4({\bf X},{\bf Y},{\bf Z})|^2 }\ .
\label{av4:op}
\eq

The integrations involved in the calculation of 
${\overline \Omega}^{(-)}_{{\bf p}}({\bf X})$
have been carried out using Monte Carlo configurations sampled from the 
probability distribution associated with the $^4$He ground state wave function 
of ref.\ci{SPW}.

\subsection{Scattering operator}

Within standard nonrelativistic nuclear
many-body theory, i.e. treating the nucleons as pointlike structureless particles,
the operator $\Gamma_{p}(1,j)$ appearing in eq.(\ref{omega:hea}) is a function of 
the particle positions ${\bf r}_1$ and ${\bf r}_j$ only. Choosing the $z$ axis 
along 
the direction of the eikonal trajectory (i.e. the direction of the momentum of the 
struck proton, specified by the unit vector ${\bf p}/|{\bf p}|$, the dependence 
of $\Gamma_{p}(1,j)$ upon $z_1$ and $z_j$ can be singled out writing
\bq
\Gamma_{p}(1,j) = \theta(z_j - z_1) \gamma_{p}(|{\bf b}_1-{\bf b}_j|)\ ,
\label{big:gamma}
\eq
where the step function preserves causality while $\gamma_{p}(b)$ is a function of 
the projection of the interparticle distance in the impact parameter plane 
(the $xy$ plane) which contains all the information on the dynamics of the scattering 
process. The function $\gamma_{p}(b)$ can be simply related to the 
coordinate space
$t$-matrix associated with the proton-nucleon (pN) potential $v_{ij}$, and 
written in
terms of the measured NN scattering amplitude at incident momentum $p$,
$f_{p}(k_t)$, as
\bq
\gamma_{p}(b) = - \frac{i}{2} \int \frac{d^2 k_t}{(2\pi)^2}\
   {\rm e}^{i {\bf k}_t \cdot {\bf b} }\ f_{p}(k_t)\ .
\label{little:gamma}
\eq

At large $p$, the experimental $f_{p}(k_t)$ is usually parametrized in the
form \ci{nndata}
\bq
f_{p}(k_t) = i\ \sigma_{pN}^{tot}(1 - i\alpha_{pN})
{\rm e}^{-\frac{1}{2}\frac{k_t^2}{B}}\ ,
\label{para:ampl}
\eq
where $\sigma_{pN}^{tot}$ and $\alpha_{pN}$ denote the total cross section and 
the ratio 
between the real and the imaginary part of the amplitude, respectively, while
$B$ is related to the range of the interaction. In the case of zero-range 
interaction, $B=0$ and the impact parameter dependence of $\gamma_{p}(b)$ reduces 
to a two-dimensional $\delta$-function.
    
To include CT, the internal structure of the proton must be explicitely taken into
account. According to the CT scenario, in the $\eep$ reaction at large 
$Q^2$ the electromagnetic interaction produces a compact three-quark state 
$|E \rangle$, which can be seen as a 
superposition of many hadronic states $| \alpha \rangle$, $| \beta \rangle \ldots$. 
This state then propagates through the nuclear medium undergoing rescattering 
processes that eventually lead to the emergence of the detected proton.
The rescattering processes can be either {\it diagonal}, when the hadronic
state $| \alpha \rangle$ does not change, or {\it off diagonal}, when a 
transition to a different state $| \beta \rangle$ is induced. The transparency
effect, i.e. the disappearance of nuclear absorption, folows from the 
cancelation between the contributions of {\it diagonal} 
and {\it off diagonal} processes at 
asymptotically high $Q^2$.

From the above discussion , it follows that, in order to describe the transition 
of FSI effects to the CT regime, one has to introduce a scattering operator acting 
in the space of the hadronic states. Its matrix element between states 
$| \beta \rangle$ and $| \alpha \rangle$, of mass $m_\beta$ and $m_\alpha$, 
respectively,  can be defined as \ci{eep} 
\bq
\langle \beta | \Gamma_p(1,j) | \alpha \rangle = \theta(z_j - z_1)
{\rm e}^{i k_{\alpha \beta} z_j }
 \gamma_{p}^{\alpha \beta}(|{\bf b}_1-{\bf b}_j|)\ ,
\label{gamma:op}
\eq
where
\bq
k_{\alpha \beta} = \frac{m_\alpha^2 - m_\beta^2}{2 E_p}\ ,
\eq
$E_p=T_p+m$ being the energy of the detected proton in the laboratory frame.
 The onset oc CT is driven by the oscillating factors 
exp($i k_{\alpha \beta} z_j$), taking into account the longitudinal momentum 
transfer associated with each transition $\alpha$ + N $\rightarrow \beta$ + N.
In analogy to eq.(\ref{little:gamma}), 
$\gamma_{p}^{\alpha \beta}(b)$ is written in terms of the amplitude of the 
process $\alpha + N \rightarrow \beta + N $:
\bq
\gamma_{p}^{\alpha \beta}(b) = - \frac{i}{2} \int \frac{d^2 k_t}{(2\pi)^2}\
   {\rm e}^{ i {\bf k}_t \cdot {\bf b} }\ 
f_{p}^{\alpha \beta}(k_t) \ ,
\label{trans:ampl}
\eq
with
\bq
f_{p}^{\alpha \beta}(k_t) = i\ \langle \beta | {\widehat \sigma} | \alpha \rangle 
(1 - i\alpha_{\alpha \beta}) 
{\rm e}^{ -\frac{1}{2} \frac{k_t^2}{ B_{\alpha \beta} } }\ .
\label{para:transampl}
\eq
In the above equation, $\alpha_{\alpha \beta}$ and $B_{\alpha \beta}$ are the 
generalization of the parameters $\alpha_{pN}$ and $B$ of eq.(\ref{para:ampl}), 
while the operator ${\widehat \sigma}$ describes the hadronic cross section. 
Unfortunately, $\alpha_{\alpha \beta}$ and $B_{\alpha \beta}$ are not known
experimentally. In our numerical calculations we have made the assumption that
the interactions responsible for off diagonal rescatterings have zero range, 
i.e. that $B_{\alpha \beta}=0$ for any $\alpha \neq \beta$. The values of
$\alpha_{\alpha \beta}$ have been varied within a reasonable range to gauge 
the sensitivity of our approach to these parameters. The results will be
discussed in the next section.

Following ref.\ci{asy}, $\langle \beta | {\widehat \sigma} | \alpha \rangle$
has been evaluated in configuration space, using
\bq
\langle \beta | {\widehat \sigma} | \alpha \rangle = 
\int d\xi d^2\rho \ \psi_\beta^\ast(\xi,\rho) \sigma(\rho) 
\psi_\alpha(\xi,\rho)\ ,
\label{matel:sigma}
\eq
where $\psi_\alpha$ and $\psi_\beta$ are harmonic oscillator wave functions
describing a quark-diquark system with longitudinal and
transverse coordinates $\xi$ and $\rho$, respectively. 
The quark-diquark oscillation frequency has been chosen to be 
$\omega_0 =$ 0.35 GeV \cite{asy}, yielding a realistic mass spectrum of the
proton excited states, while 
$\sigma(\rho)$ has been parametrized in the form \ci{sigma}
\bq
\sigma(\rho) = \sigma_0 \left[ 1 - {\rm e}^{-\left(\frac{\rho}{\rho_0}\right)^2} 
\right]\ ,
\label{def:sigma}
\eq
with $\sigma_0 = 2 \sigma_{pN}$ and $\rho_0$ adjusted in such a way as to 
reproduce the experimental pN total cross section. 

A scattering operator suitable to describe the onset of CT, denoted 
$\Omega_{CT}(R)$, can be constructed using eq.(\ref{omega:hea}) and 
the two-body scattering operators $\Gamma_p(1,j)$ whose matrix elements are 
defined by eq.(\ref{gamma:op}): 
\be
\nonumber
\Omega_{\bf p}^{CT}(R) & = & \frac { P_z \langle p | 
\prod_{j=2}^A \left[ 1 - \Gamma_{p}(1,j) \right] | E \rangle }{
 \langle p | E \rangle }\\
\nonumber 
               & = & 1 - P_z \sum_{j=2}^A\ \sum_{\alpha} 
\langle p | \Gamma_p(1,j) | \alpha \rangle \frac{ 
\langle \alpha | E \rangle }{ \langle p | E \rangle }\\
               & + &  P_z \sum_{k>j=2}^A\ \sum_{\alpha\beta}
\langle p | \Gamma_p(1,k) | \beta \rangle
\langle \beta | \Gamma_p(1,j) | \alpha \rangle 
\frac{ \langle \alpha | E \rangle }{ \langle p | E \rangle }
 + \ldots\ ,
\label{omega:ct}
\ee
where $| p \rangle$ is the state describing the detected 
proton. For the compact state $| E \rangle$ produced at the electromagnetic
vertex, we have used the same configuration space wave function employed in 
refs.\ci{asy,eep}
\bq
\langle \rho | E \rangle \propto {\rm e}^{-C \rho^2 Q^2}\ ,
\label{ejectile:wf}
\eq
with $C=1$. It has to be pointed out that, as shown in ref.\ci{asy}, the 
missing momentum distribution
is not sensitive to the choice of $C$ as long as $C \ge$ 0.05.

%%%%%%%%%%%%%%%%%%%%%%%%%%%%%%%%%%%%%%%%%%%%%%%
\section{Results}
%%%%%%%%%%%%%%%%%%%%%%%%%%%%%%%%%%%%%%%%%%%%%%%

Using the functions $\chi_0({\bf X})$ and 
${\overline \Omega}^{(-)}_{{\bf p}}({\bf X})$ 
defined by
eqs.(\ref{chi:zero}) and (\ref{av4:op}), respectively, the missing 
momentum distribution associated with the $^{4}He(e,e^\prime p)^{3}H$ process, 
denoted $W_{{\bf p}}({\bf p}_m)$, can be readily obtained from
\bq
W_{{\bf p}}({\bf p}_m) = \left| \int d^3X {\rm e}^{i{\bf p}_m \cdot {\bf X}}
\chi_0({\bf X}) {\overline \Omega}^{(-)}_{{\bf p}}({\bf X}) \right|^2\ .
\label{mom:dist}
\eq
The results discussed in the present paper have been obtained using the
overlap $\chi_0({\bf X})$ computed in ref.\ci{SPW} using the 
Argonne v14 two-nucleon interaction and the Urbana VII three-body 
potential. The scattering operator 
${\overline \Omega}^{(-)}_{{\bf p}}({\bf X})$
has been calculated carrying out the integrations involved in 
eq.(\ref{av4:op}) with the Monte Carlo method, using a configuration
set sampled from the probability distribution associated with the 
$^4$He wave function of ref.\ci{SPW}. 

\begin{figure}
\begin{center}
\leavevmode
\centerline{\epsfig{figure=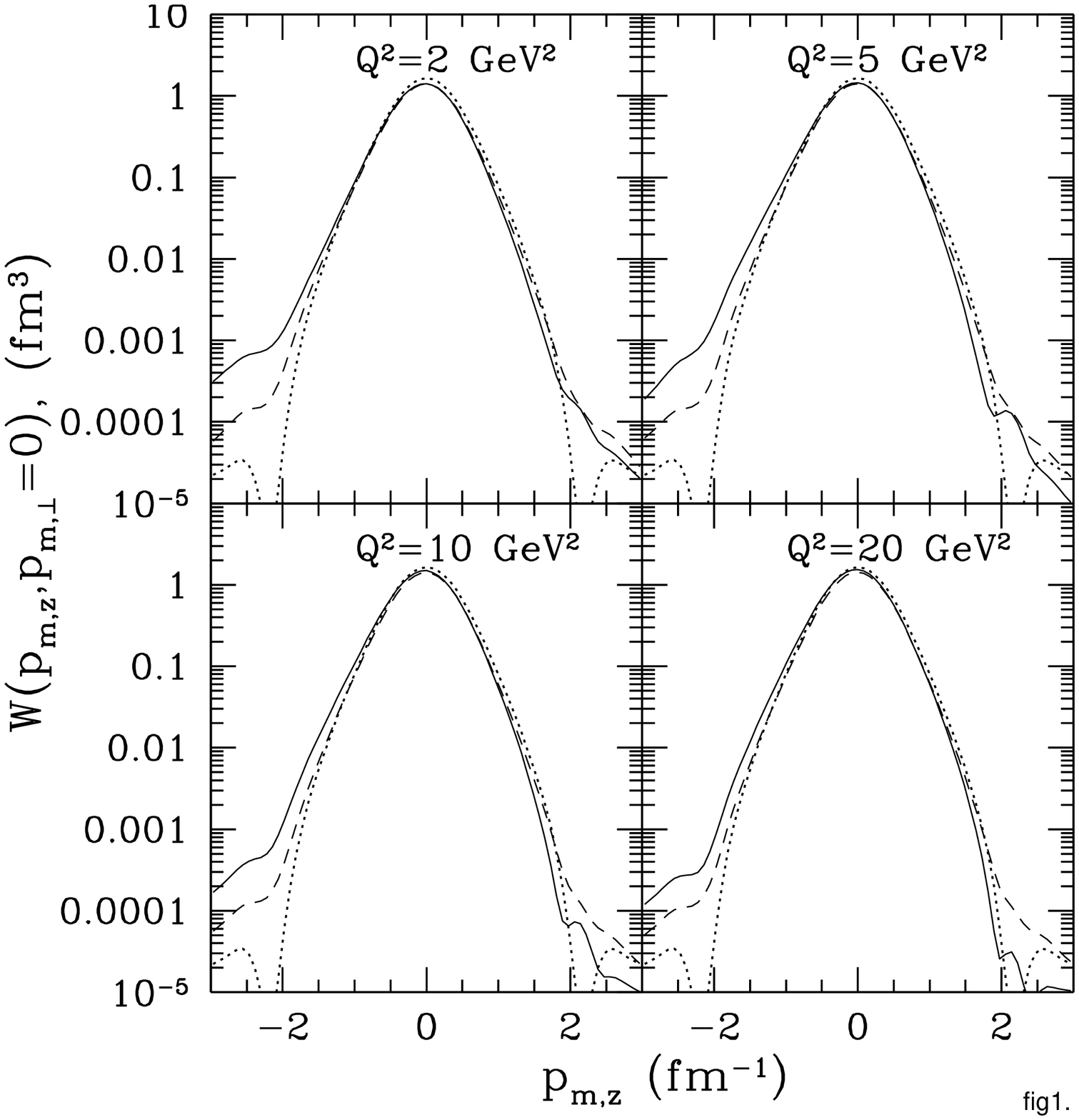,width=10.5cm}}
\caption{
Missing momentum distribution for the process $^4He(e,e^\prime p)^3H$,
evaluated at the top of the quasifree peak in parallel kinematics.
The dotted line shows the PWIA result, whereas
the dashed and solid curves correspond to the calculations including 
FSI effects with and without color transparency, respectively.        
}
\label{f1}
\end{center}
\end{figure}

\begin{figure}
\begin{center}
\leavevmode
\centerline{\epsfig{figure=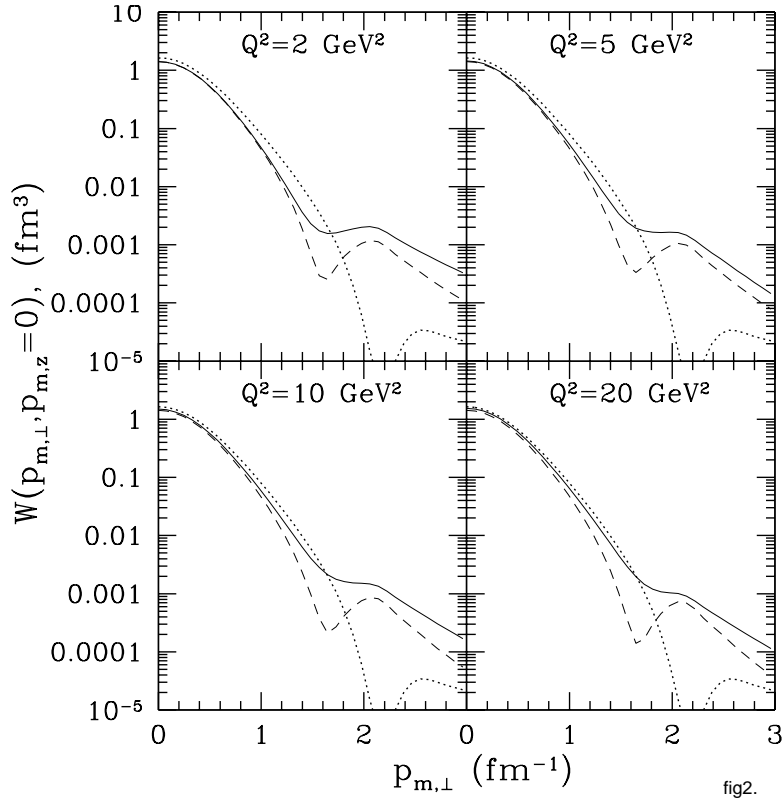,width=10.5cm}}
\caption{
Missing momentum distribution for the process $^4He(e,e^\prime p)^3H$,
evaluated at the top of the quasifree peak in perpendicular kinematics.
The dotted line shows the PWIA result, whereas
the dashed and solid curves correspond to the calculations including
FSI effects with and without color transparency, respectively.
}
\label{f2}
\end{center}
\end{figure}

In figs. \ref{f1} and \ref{f2} we show $W_{{\bf p}}({\bf p}_m)$ evaluated 
at the top 
of the quasi free peak, i.e. at $\omega = Q^2/2m$, for parallel
($p_{m,\perp} = |{\bf p}_m \times {\bf q}| = 0$) and 
perpendicular ($p_{m,z} = |{\bf p}_m \cdot {\bf q}| = 0$) 
kinematics, respectively. 
Each figure
has four panels, corresponding to different values of $Q^2$ ranging from 
2 (GeV/c)$^2$ to 20 (GeV/c)$^2$. The dotted line shows the PWIA 
(i.e. ${\overline \Omega}^{(-)}_{{\bf p}}({\bf X}) \equiv 1$) result, whereas 
the dashed and solid curves correspond to the calculations including FSI 
effects with and without CT, respectively. 
The configuration set employed in our calculations allows for an accurate
determination of the missing momentum distribution over a large momentum
range. However, in the region where $W_{{\bf p}}$ becomes very small
($ \le 10^{-4}$ fm$^3$), the statistical uncertainty of the Monte Carlo 
calculation becomes sizeable. 
Although our results indicate that the first two-three excited states 
saturate the contribution of the off-diagonal rescatterings 
at missing momentum less than 300 MeV/c, we have included six intermediate 
states in all numerical calculations.

Fig. \ref{f1} shows that, while within PWIA 
$W_{{\bf p}}(p_{m,z}) = W_{{\bf p}}(-p_{m,z})$, FSI produce a 
forward bakward asymmetry, whose origin has to be ascribed to 
the effect of the real part of the NN scattering amplitude
\ci{bi2} and to the fact that the cancellation between the contributions
of diagonal and off-diagonal rescattering processes depends upon 
the value of the missing momentum \ci{NNNCT}. 

The main features of the distorted missing momentum 
distribution are the quenching in the region of the maximum, 
corresponding to $p_{m,z} \sim$ 0, and the enhancement of the tail at 
negative $p_{m,z}$. As expected, inclusion of CT reduces the effect of FSI. 

A more complicated structure is observed in the case of 
perpendicular kinematics, shown in fig.\ref{f2}. 
 The mimimum displayed by the PWIA missing momentum distribution, 
almost completely washed out by FSI, reappears at lower values of 
$p_{m,\perp}$ when the effect of CT is included.
Note that at large $p_{m,\perp}$ ($p_{m,\perp} \ge 2$ fm$^-1$) 
the momentum distribution is dominated by FSI and that the 
inclusion of CT results in a sizable suppression.

The complex pattern of quenching and
enhancement of the PWIA $W_{{\bf p}}$ has to be ascribed to the combined
effects of FSI and strong NN correlations in the initial state.
A similar behavior has been found in ref.\ci{bi2}, where the
semi-inclusive process $^4He(e,e^\prime p)X$ has been analyzed
using the somewhat simplified Jastrow model to describe 
NN correlations. More recently, the $^4He(e,e^\prime p)X$ reaction 
has been studied using a four-body wave function including 
noncentral correlations induced by the tensor component of the 
 NN interaction \ci{ctm}. The distorted momentum distributions 
of refs.\ci{bi2} and \ci{ctm} exhibit the same pattern of FSI 
effects on the S-wave contribution, while the D-wave is only 
weakly distorted. The main difference between the two approaches 
cancellation between the effects of central and tensor correlations
at large missing momenta, leading to a suppression of the distortion 
by a factor of $\sim$ 2.

The mimimum displayed by the PWIA missing momentum distribution, almost 
completely washed out by FSI, reappears at lower values of 
$p_{m,\perp}$ when the effect of CT is included.
Figs. \ref{f1} and \ref{f2} show that 
at $Q^2$ = 2 (GeV/c)$^2$ and in absence of CT the 
missing momentum distribution at $p_{m,\perp} = p_{m,z} =$ 0
 gets quenched by 18 $\%$ on account of FSI.

\begin{figure}
\begin{center}
\leavevmode
\centerline{\epsfig{figure=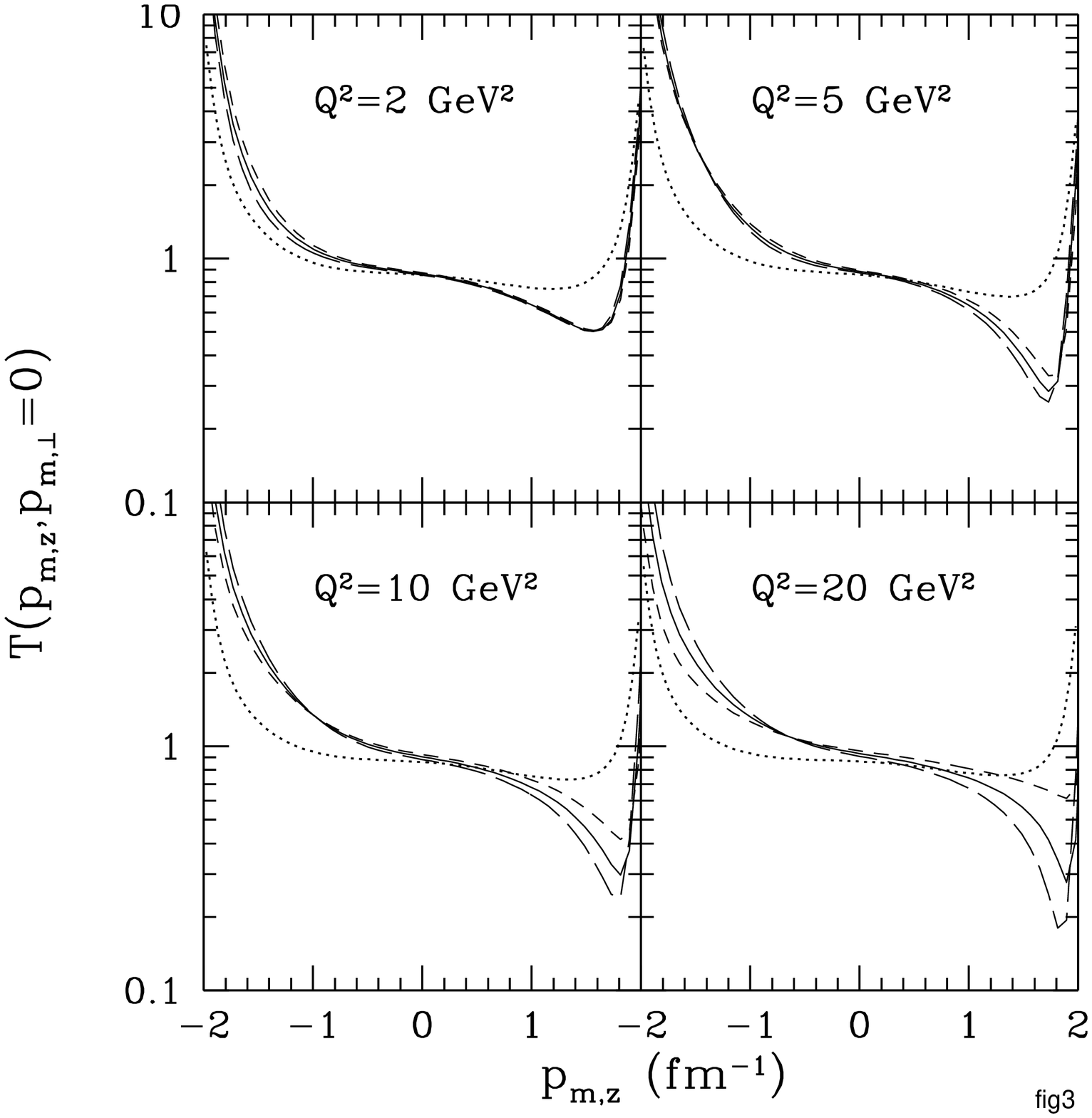,width=10.5cm}}
\caption{
Transparency ratios defined by eq.(\protect{\ref{loc:transp}}) and
corresponding to the missing momentum distributions of 
fig.\protect\ref{f1}.
}
\label{f3}
\end{center}
\end{figure}

\begin{figure}
\begin{center}
\leavevmode
\centerline{\epsfig{figure=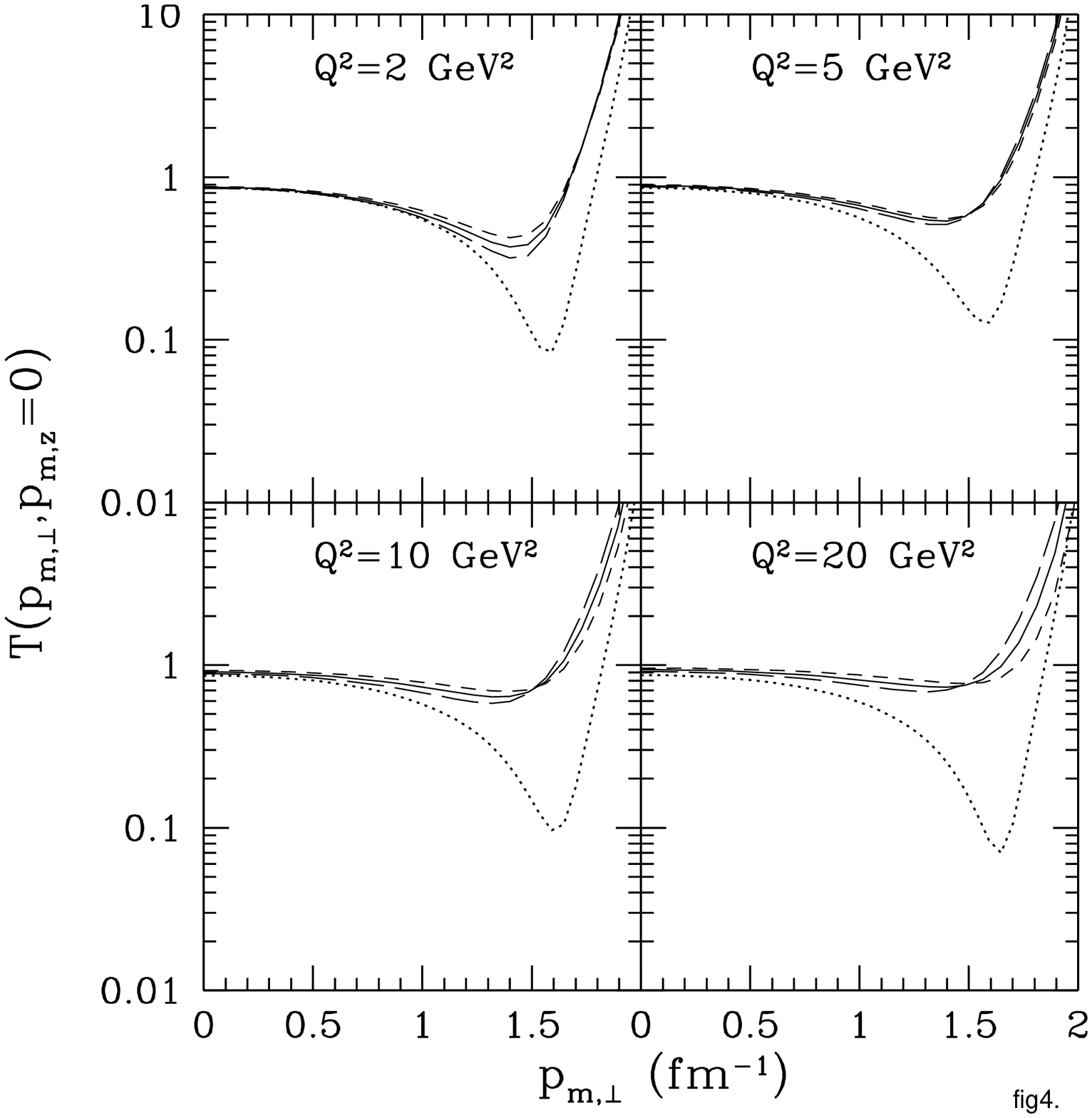,width=10.5cm}}
\caption{
Transparency ratios defined by eq.(\protect{\ref{loc:transp}}) and
corresponding to the missing momentum distributions of
fig.\protect\ref{f2}.
}
\label{f4}
\end{center}
\end{figure}

The effects of FSI can be best observed in the ratio
\bq
T_p({\bf p}_m) = \frac{W_{{\bf p}}({\bf p}_m)}
{\left| \int d^3X {\rm e}^{i{\bf p}_m \cdot {\bf X}} \chi_0({\bf X}) 
\right|^2}\ ,
\label{loc:transp}
\eq
shown in figs. \ref{f3} and \ref{f4}. It clearly follows from the 
definition that within PWIA $T_p({\bf p}_m)\equiv 1$. 
When FSI are included, $T_p({\bf p}_m)$ is a function of the 
missing momentum and can take values both above and below than 
unity, reflecting the fact that the distorted missing momentum 
distribution can be larger or smaller than the PWIA momentum
distribution. Hence, in spite of the analogy between 
eq.(\ref{loc:transp}) and the definition of the nuclear transparency, 
the quantity 1 - $T_p({\bf p}_m)$ cannot be simply interpreted as
 the nuclear absorption experienced by a proton carrying momentum 
${\bf p}$. 

The calculated $T_p$ corresponding to parallel and perpendicular
kinematics are presented in figs. \ref{f3} and \ref{f4}, 
respectively. The dotted curve shows the results obtained 
when no off-daigonal rescatterings are included, i.e. in absence of CT. 
The solid, dashed and long-dashed
curves correspond to calculations in which CT effects have been taken 
into account by properly including off-diagonal rescatterings 
and using three different sets of parameters $(\alpha_1,\alpha_2)$ 
to describe the real part of the scattering amplitudes associated 
with diagonal ($\alpha + N \rightarrow \alpha + N$, with $\alpha \ne p$)
and off-diagonal ($\alpha + N \rightarrow \beta + N$, with 
$\alpha \ne \beta$) processes.
These amplitudes have been parametrized according to:
\be
\nonumber
Re\ f(\alpha + N \rightarrow \alpha + N) & = & 
\alpha_1 Re\ f(p + N \rightarrow p + N) \\
 & = & \frac{\alpha_1}{3}
\left[ \alpha_{pp}\sigma_{pp}+2\alpha_{pn}\sigma_{pn} \right]\ ,
\ee
and
\bq
Re\ f(\alpha + N \rightarrow \beta + N)= \alpha_2\  
Im\ f(\alpha + N \rightarrow \beta + N)\ .
\eq
The results of our calculations turn out to be insensitive to the 
value of $\alpha_1$, so we set $\alpha_1 = 1$ and show the effect of 
varying $\alpha_2$ only. The solid, dashed and long dashed curves
of Fig. 3 correspond to the sets $(\alpha_1,\alpha_2)$ = (1,0),(1,0.5) and
(1,-0.5). In parallel kinematics the dependence upon 
$\alpha_2$ appears to be sizable, particularly at the largest value of $Q^2$. 
On the other hand, Fig. \ref{f4} shows that in perpendicular kinematics, where 
the effect of CT at large missing momentum is large, the theoretical 
error bar associated with the uncertainty in $\alpha_2$ is rather small.

\begin{figure}
\begin{center}
\leavevmode
\centerline{\epsfig{figure=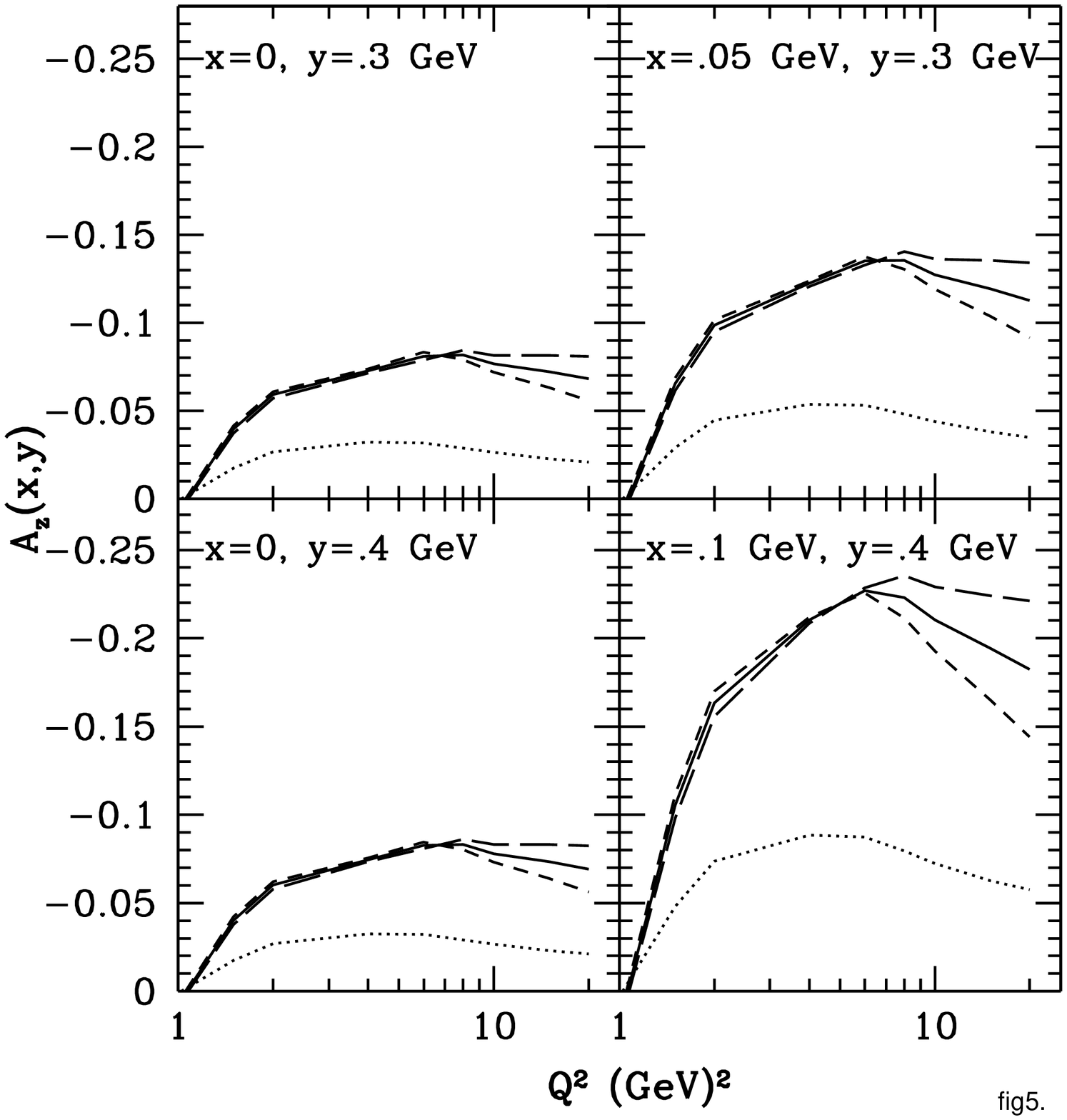,width=10.5cm}}
\caption{
$Q^2$ dependence of the longitudinal forward-backward asymmetry, defined as in 
eqs.(\protect\ref{asy:1})-(\protect\ref{asy:2}), corresponding to four 
different kinematical windows.
}
\label{f5}
\end{center}
\end{figure}

In fig. \ref{f5}, we present the longitudinal forward-backward asymmetry of
the missing momentum distribution defined as 
\bq
A_z(x,y)=\frac{N_+-N_-}{N_++N_-}\ ,
\label{asy:1}
\eq
where
\bq 
N_{\pm} = \int_{\pm x}^{\pm y} dp_{m_z}\ 
W_{{\bf p}}\left(p_{m_z},p_\perp=0\right)\ .
\label{asy:2}
\eq
We have evaluated $A_z$ from the above equations for four kinematical windows: 
$(x,y)$ = (0,0.3), (0,0.4), (0.05,0.3) and (0.1,0.4) GeV/c. To illustrate the 
contribution of the off-diagonal rescattering processes to $A_z$, the results 
obtained setting these contributions to zero are shown by the dotted line. 
As in fig. \ref{f3} and \ref{f4}, the solid, dashed and long-dashed lines 
correspond to different choices of the parameter $\alpha_2$.
The results of fig. \ref{f5} show that even at moderate 
$Q^2 \sim 2.5$ (GeV/c)$^2$, i.e. 
in the region relevant to the $(e,e^\prime p)$
experimental program at the Thomas Jefferson National Accelerator Facility
(TJNAF), off-diagonal rescatterings are responsible for
more than 60 $\%$ of the calculated asymmetry for all of considered kinematical 
windows.
In a previous study of the asymmetry in semi-inclusive $(e,e^\prime p)$ processes
 we have found a contribution of 15-20 $\%$ and 10-15 $\%$ in the case of 
$^{16}O$ and $^{40}Ca$, respectively \ci{asy}. The results of the present 
calculation confirm our conclusion that CT effects are larger 
in light nuclei.
At $Q^2 le 20$ (GeV/c)$^2$, our calculations show a weak dependence of $A_z$
upon the value of the parameter $\alpha_2$. Hence, the uncertainty associated 
with the chioce of $\alpha_2$ does not prevent one from extracting an 
unambiguous signature of the onset of CT from the asymmetry.
 
\section{Summary and conclusions}

We have carried out a calculation of the missing momentum distribution of the 
process $^4$He$(e,e^\prime p)^3$H in quasielastic kinematics, i.e. at 
$\omega \sim Q^2/2m$, in the range $2 \le Q^2 \le 20$ (GeV/c)$^2$. 

The PWIA momentum distribution, evaluated
using highly realistic many-body wave functions and the Monte Carlo method, 
has been corrected to include the effects of FSI, treated within a 
coupled-channel multiple scattering approach suitable to describe the
possible onset of CT. It has to be emphasized that our approach allows 
for a consistent treatment of short range NN correlations in both the initial
and final state.

Inclusion of FSI leads to the appearance of a complex pattern of distorsions
of the PWIA momentum distribution, both in parallel and perpendicular 
kinematics. 
Sizable CT effects are observed in perpendicular kinematics
at $p_{m \perp} \sim 1.5$ fm$^{-1}$ over the whole $Q^2$ range. 
The CT signal turns out to be much larger than the theoretical 
uncertainty associated with the parametrization of the amplitudes
for off-diagonal rescattering. 
The calculated forward-backward asymmetry also shows a significant
CT effect already at $Q^2 \sim$ 2.5 (GeV/c)$^2$. 

In conclusion, our results seem to indicate that the experimental study 
of the exclusive 
channels in $(e,e^\prime p)$ processes off few-nucleon system may 
give a clue to the issue of the possible manifestation of CT 
 in the domain of moderate $Q^2$ ($\le$ 3 (GeV/c)$^2$), covered by 
the existing electron scattering facilities. 

\begin{ack}

It is a pleasure to thank Rocco Schiavilla and Robert B. Wiringa for
providing Monte Carlo configurations of the $^4$He ground state and the
results of their calculation of the $\langle ^3H | ^4He \rangle$ overlap.
One of the authors (AAU) gratefully acknowledges the hospitality provided
by the Sezione Sanit\`a of the Italian National Institute for Nuclear
Research (INFN) and the Institut f\"{u}r Kernphysik, Forschungszentrum
J\"{u}lich, where part of the work described in this paper has been carried
out.

\end{ack}

\clearpage

%%%%%%%%%%%%%%%%%%%%%%%%%%%%%%%%%%%%%%%%%%%%%%%%%%%%%%%%%%%%%%%%%%%%%%%%%
%%%%%%%%%%%%%%%%%%%%%%%%%%%%%%%%%%%%%%%%%%%%%%%%%%%%%%%%%%%%%%%%%%%%%%%%%%

\clearpage

%%%%%%%%%%%%%%%%%%%%%%%%%%%%%%%%%%%%%%%%%%%%%%%%%%%%%%%%%%%%%%%%%%%%%%%%%
%%%%%%%%%%%%%%%%%%%%%%%%%%%%%%%%%%%%%%%%%%%%%%%%%%%%%%%%%%%%%%%%%%%%%%%%%%

\clearpage

%%%%%%%%%%%%%%%%%%%%%%%%%%%%%%%%%%%%%%%%%%%%%%%%%%%%%%%%%%%%%%%%%%%%%%%%%
%%%%%%%%%%%%%%%%%%%%%%%%%%%%%%%%%%%%%%%%%%%%%%%%%%%%%%%%%%%%%%%%%%%%%%%%%%

\clearpage

%%%%%%%%%%%%%%%%%%%%%%%%%%%%%%%%%%%%%%%%%%%%%%%%%%%%%%%%%%%%%%%%%%%%%%%%%
%%%%%%%%%%%%%%%%%%%%%%%%%%%%%%%%%%%%%%%%%%%%%%%%%%%%%%%%%%%%%%%%%%%%%%%%%%

\clearpage

%%%%%%%%%%%%%%%%%%%%%%%%%%%%%%%%%%%%%%%%%%%%%%%%%%%%%%%%%%%%%%%%%%%%%%%%%
%%%%%%%%%%%%%%%%%%%%%%%%%%%%%%%%%%%%%%%%%%%%%%%%%%%%%%%%%%%%%%%%%%%%%%%%%%

\end{document}